\documentclass[conference]{IEEEtran}
\usepackage[a4paper,top=0.81in, left=0.575in, right=0.575in, bottom=0.81in]{geometry}
\IEEEoverridecommandlockouts
\usepackage{amsmath,amssymb,amsfonts}
\usepackage[ruled,vlined]{algorithm2e}
\usepackage{algorithmic}
\usepackage{graphicx}
\usepackage{textcomp}
\usepackage{todonotes}
\usepackage{xcolor}
\usepackage{cite}
\usepackage{mathtools}
\usepackage{mathabx}

\ifCLASSOPTIONcompsoc
    \usepackage[caption=false, font=normalsize, labelfont=sf, textfont=sf]{subfig}
\else
\usepackage[caption=false, font=footnotesize]{subfig}
\fi

\usepackage[nolist,nohyperlinks]{acronym}

\begin{acronym}
\acro{MIMO}{Multiple-input Multiple-output}
\acro{LIS}{Large Intelligent Surface}
\acro{DL}{Deep Learning}
\acro{CSI}{Channel State Information}
\acro{5G}{5th generation of wireless networks}
\acro{6G}{6th generation of wireless networks}
\acro{SNR}{Signal-to-Noise Ratio}
\acro{ML}{Machine Learning}
\acro{DAE}{denoising autoencoder network}
\acro{RSS}{Received Signal Strength}
\acro{RTI}{Radio Tomographic Image}
\acro{WSN}{Wireless Sensor Network}
\acro{ECDF}{Empirical Cumulative Distribution Function}
\acro{LoS}{Line of Sight}
\acro{ToF}{Time of Flight}
\acro{FMCW}{Frequency Modulated Carrier Wave}
\acro{MF}{Matched Filter}
\acro{USRP}{Universal Software Radio Peripheral}
\end{acronym}

\begin{document}

\title{Radio Sensing with \\ Large Intelligent Surface for 6G
\thanks{This work has been submitted to IEEE for possible publication. Copyright may be transferred without notice, after which this version may no longer be accessible. 
This project has received funding from the European Union’s Horizon 2020 research and innovation programme under the Marie Skłodowska-Curie Grant agreement No. 813999.}
}
\author{Cristian J. Vaca-Rubio$^{1,*}$, Pablo Ramirez-Espinosa$^1$, Kimmo Kansanen$^2$, \\ Zheng-Hua Tan$^1$ and Elisabeth de Carvalho$^1$ \\
$^1$Department of Electronic Systems, Aalborg University, Aalborg, Denmark \\
$^2$Norwegian University of Science and Technology, Trondheim, Norway\\
\{cjvr$^*$, pres, zt, edc\}@es.aau.dk, kimmo.kansanen@ntnu.no \\
$^*$Corresponding Author
}

\maketitle

\begin{abstract}
This paper leverages the potential of Large Intelligent Surfaces (LIS) for radio sensing in 6G wireless networks. Major research has been undergone about its communication capabilities but it can be exploited as a formidable tool for radio sensing. By taking advantage of arbitrary communication signals occurring in the scenario, we apply direct processing to the output signal from the LIS to obtain a radio map that describes the physical presence of passive devices (scatterers, humans) which act as virtual sources due to the communication signal reflections. We then assess the usage of machine learning (k-means clustering), image processing and computer vision (template matching and component labeling) to extract meaningful information from these radio maps. As an exemplary use case, we evaluate this method for both active and passive user detection in an indoor setting. The results show that the presented method has high application potential as we are able to detect around 98\% of humans passively and 100\% active users %\pres{I think these are the results for the worst case right? Maybe it's not good to show this case in the abstract, since may condition the reader letting him/her think that, as best, only 80\% of the users are detected} 
by just using communication signals of commodity devices even in quite unfavorable Signal-to-Noise Ratio (SNR) conditions. 
\end{abstract}

\section{Introduction}
\label{Introduction}
%\pres{Since we are running out of space, I suggest shortening the intro}

%Throughout the history, every generation for wireless systems created new services transforming in this way people daily lives.
%In the \ac{2G}, it started the mobile voice services. Then, in the \ac{3G} mobile data services appeared. Next, the \ac{4G} enabled the Mobile Broadband services (MBB) and 
%\ac{5G} surpassed the constraints of the \ac{4G} by addressing \ac{eMBB}, \ac{URLLC}  and \ac{mMTC} services. However, 
Services offered by the \ac{5G} are expected to be further enhanced. %\pres{For me, this historical overview of the mobile generations is a candidate to be shortened/removed}. 
For that purpose, %it will likely move from mmWaves up to the THz and sub-THz frontier \cite{rappaport2019wireless}. The THz and mmWave applications coexisting with the sub-mmWave bands will be one of the main potentials for 6G \cite{rappaport2019wireless, chaccour2021seven}. As a consequence, working in higher frequencies and exploiting greater bandwitdh leads to a dominant \ac{LoS} propagation \cite{do2020terahertz} while empirical pathloss models show that even in non-\ac{LoS} conditions there is one or few dominant paths \cite{xing2021propagation}. To compensate this high frequency pathloss, 
sensing and positioning based on radio signals \cite{zhang2017indoor, chiou2009design} might play a vital role in combination with a proper beam alignment \cite{song2019efficient, noh2017multi}. So, what will be the new features of the \ac{6G}? We envisage that one of the main features of 6G will be integrating sensing and  radio-based imaging, providing context information which improves the ultimate goal of communication. It will enable
merging communications and new applications such as 3D imaging and sensing \cite{latva2019key}. %Totally new services such as telepresence and mixed reality will be possible thanks to high resolution imaging
%and sensing, accurate positioning, wearable displays, mobile robots and drones, specialized processors, and next-generation wireless networks.

Sensing can be regarded as the ability of wireless systems to process the signals with the aim of describing the physical environment. There are different methodologies to perform sensing using wireless signals. Essentially, some of these methods use dedicated signals and/or specific hardware \cite{adib2014real,stasiak2017fmcw, wilson2010radio,lee2019variational}, while others use  communication signals of commodity devices to perform the sensing task \cite{wang2016rt,pu2013whole}. As an example of the first type, in \cite{wilson2010radio, lee2019variational} they employ \ac{RTI}, which is a \ac{RSS}-based technology for rendering physical objects in wireless networks. They create a radio map based on the \ac{RSS} variations due to objects presence in the scenario by deploying nodes around the room conforming a \ac{WSN}.  In turn, by making use of the communication signals occurring in an environment and avoiding dedicated transmissions \cite{wang2016rt, pu2013whole}, one can rely on properties of the wireless channel such as the \ac{CSI} using commodity Wi-Fi devices, to perform sensing tasks as human gesture recognition or fall detection. Works like the ones presented in \cite{adib2014real, stasiak2017fmcw} capture the reflections of wireless signals, similar to the radar principle. They use Wi-Fi devices and to estimate the \ac{ToF} they need to rely on the \ac{FMCW} technique. This enables applications such as human pose estimation and breath monitoring. When relying on a radar-like approach,  high accuracy is obtained, but needs the addition of customized extra hardware (such as \ac{USRP} \cite{stasiak2017fmcw}) to generate the signal that sweeps the frequency across time. %\ac{6G} will allow the convergence between intelligent wireless sensing and communications \cite{latva2019key}, being AI a powerful solution for this convergence \cite{letaief2019roadmap}. %\pres{If we need more space, this paragraph can be shortened a little bit too}

In the context of communications, the \ac{MIMO} technique is a fundamental technology in \ac{5G} with the main purpose of increasing area spectral efficiency \cite{larsson2014massive}. In massive \ac{MIMO}, the base station is equipped with a very large number of antennas. Intending to push their benefits to the limit and look towards post-5G, researchers are defining a new generation of base stations that are equipped with an even larger number of antennas. The concept of \ac{LIS} designates a large continuous electromagnetic surface able to transmit and receive radio waves. %\pres{I would jump here directly to the popularity in the research world of LISs, and that they appear as a "hypothetical and promising" extensions of mMIMO} 
%These large surfaces can be placed on walls/ceilings and are easily integrable into the surroundings. 
In practice, an \ac{LIS} is composed of a collection of closely spaced tiny antenna elements. While the potential for communications of \ac{LIS} is being investigated, these devices offer possibilities that are not being understudied accurately, i.e., environmental sensing based on radio images \cite{vaca2021assessing}. %Having a huge array of antennas is really advantageous for extremely directional applications. When frequency increases, the spatial resolution becomes finer allowing then the radio map generation that can allow to interpret the channel characteristics. Indeed, such large surfaces contain many antennas that can be used as sensors of the environment based on the \ac{CSI}. This can be useful for improving the allignment of directional antennas, provide on-the-fly localization and a real time adaptation of the wireless capabilities. 

%The large aperture and high number of antenna elements of \ac{LIS} can be used for performing an accurate environment sensing while providing a huge amount high-dimensional data. In this way, \ac{ML} and deep learning are useful for understanding radio environmental maps. These radio maps can be mappings of the signals into an image structure, describing the propagation environment. The input to this deep learning networks can be addressed from different perspectives: channel state information (CSI) or reconstruction of radio environmental maps to exploit convolutional neural networks. %The output can be several parameters depending on the optimization problem we are trying to solve. %For example, with a radio environmental image, computer vision approaches can be exploited for tracking users and performing any strategy for bitrate optimization. Because the network can be trained offline, one of the key benefits of this method is that we can improve the LIS performance in a real time fashion.

%Currently, wireless networks exploit radio waves for communications mostly. However, another advantageous way of using radio signals is for sensing the environment. The main goal of sensing is taking advantage of the radio propagation environment in order to describe the physical phenomena in a specific scenario.
Due to the increasing interest in both sensing and \ac{LIS}, and motivated by their future integration in communication systems, in this work, we are focusing on \ac{LIS} sensing capabilities. Specifically, the contributions of this work can be summarized as follows: %\pres{Put some bridge between all the literature review and the contributions. E.g., due to the increasing interesting in both sensing and LIS, and motivated by their future integration in communication systems, we...}
\begin{itemize}
    \item We present a method that enables reconstructing a radio map of the propagation environment using an indoor \ac{LIS} deployment in the ceiling. This radio map shows the presence of active and passive (scatterers/humans) users in the environment by relying on environmental communication signals. %\KK{What is a radio map in this work, define the concept precisely. Here, or elsewhere. Make a distinction to earlier methods, if you can.}
    \item We solve a problem of active and passive multi-user detection in the scenario using the reconstructed radio maps. The solution is based on the k-means clustering of the radio maps, followed by the application of image processing to enhance the quality and computer vision to perform the detection. %\KK{Remove "to evaluate...". Just claim "Detection" of persons, not tracking. Add a sentence summarizing the solution strategy.}
\end{itemize}
%Detecting active users can be useful, for instance, in robot tracking in factories. These robots are sending signals for their coordination and we can take advantage of them to monitor failures or automatic supervision of their tasks.
Detecting passive users is of great interest as we are relying on environmental radio signals and do not need dedicated devices. This could be quite useful for human intrusion detection  or to optimize beamforming towards the passive human enabling the access phase with an optimized radiation pattern. 
These radio maps could be used for further tasks related to the environmental description and location of scatterers. Hence,  we  analyze  the  feasibility  of  this proposal  to  determine the precision in users detection and environmental description by exploiting the information provided by the radio propagation environment as a radio map. We measure the detection accuracy as the number of users detected while also verifying the positioning accuracy. Furthermore, this work leverages the potential of the combination between wireless sensing, \ac{LIS} and computer vision for \ac{6G}.  %\KK{We precisely do NOT want to determine the precision of positioning. We measure the detection accuracy by verifying that the detected object is also correctly positioned.}

\section{Problem Formulation and System Description}
%\pres{Why everything is in the problem formulation section? I would put 3 short sections: one with the system model, another one with the obtention of the radio maps (explaining MF and spherical steering vectors), and another section with the object/user detection. Finally, the cherry on top is the validation section where you introduce the scenario in one subsection and the results in other one.}
%In the last sections, we have revised the state-of-the-art for wireless sensing, provided an overview of the future \ac{6G} capabilities as well as the advantages of computer vision and image processing for understanding image information. 
%We highlighted one of the main enabling features of \ac{6G} will be radio sensing. With the aim of demonstrating the usefulness of \ac{LIS}, radio based imaging and computer vision, we here present a baseline problem in order to analyze the sensing potential of \ac{LIS}. \pres{Would remove this paragraph}

Let us consider an indoor scenario where $U_a$ users are randomly deployed in a room\footnote{Please note we refer as user to anything using a transmitter, such as robots, IoT devices or smartphones.}. We assume that all these users are commodity wireless devices that are fulfilling their communication tasks while we take advantage of them to perform the sensing. %\KK{environmental signals is a poor term, re-invite}. 
The goal is two-fold:
\begin{enumerate}
    \item Construction of radio maps based on commodity devices' communication signals. %\KK{At this point you should have decided if it is a map or an image. Choose one.}
    \item Using these radio maps to detect both $U_a$ active users and $U_p$ passive scatterers/objects and/or humans.
\end{enumerate}
%With this radio map construction, we propose a use case in which we aim to detect both active and passive humans.% Especially, passive human detection might be of great interest for exemplary use cases such as physical layer security or performing Electromagnetic Fields (EMF) control to avoid human radiation exposure. %\KK{(The previous sentence undermines the importance of the user detection. You want to emphasize the importance of person detection, instead. You can, thereafter claim some exemplary use cases like EMF control and PHY layer security.)}. 

For this task, we assume that an \ac{LIS} of $M$ antenna elements is placed along the ceiling, whose physical aperture comprises its whole area. For simplicity, we assume an ideal \ac{LIS} composed of isotropic antennas and physical effects such as mutual coupling are ignored\footnote{The mutual coupling effect is usually modeled as a coupling matrix which accounts for the effect of neighboring antennas \cite{su2001modeling}. Once the coupling matrix is estimated, its effect can be compensated. This motivated avoiding the addition of extra complexity to the model, as it would not impact on the main findings and conclusions of this work.}. The sensing problem reduces to determine, from the superposition of the received signals from each of the $U_a$ users at every of the $M$ \ac{LIS} elements, the $(x, y)$ coordinates (a.k.a the position) of the $U_a$ users involved in the scenario as well as the $U_p$ passive humans. The superposed complex baseband signal received at the \ac{LIS} is given by  
\begin{equation}
    \label{eq:RecSignal}
	\mathbf{y} = \sum_{u=1}^{U_a}\mathbf{h_u}x_u + \mathbf{n},
\end{equation}
with $x_u$ the transmitted (sensing) symbol from user $u$, $\mathbf{h_u}\in\mathbb{C}^{M\times 1}$ the channel vector from a specific position of user $u$ to each antenna-element, and $\mathbf{n}\sim\mathcal{CN}_{M}(\mathbf{0},\sigma^2\mathbf{I}_M)$ the noise vector. Please note we are considering a narrowband transmission, avoiding frequency selectivity effects. %\KK{What is the technical reason that makes this necessary, to avoid frequency selectivity effects? In such a case mention it. Indoor delay spreads are not that large.}. %In this way, we conclude this active/passive user detection is one of the applications we explore based on the radio map reconstruction method presented in this paper. \KK{Last sentence appears redundant.}

\section{LIS radio map generation}
\label{subsec:radioImage}
Due to the large physical aperture of the deployment in comparison with the distance between the transmitters and the \ac{LIS}, %\KK{(compared to the distance between transmitters and the LIS)}
spherical wave propagation needs to be taken into account. Describing the channel in (\ref{eq:RecSignal}) as the sum of $N_r$ paths, the channel from user $u$ to the $i$-th antenna element can be regarded as
\begin{equation}
    h_{u,i} = \underbrace{V_{1,i} e^{-j\frac{2\pi}{\lambda}d_{1,i}}}_{A} + \underbrace{\sum_{n=2}^{N_r} V_{n,i} e^{-j\frac{2\pi}{\lambda }d_{n,i}}}_{B}, 
\end{equation}
where $A$ and $B$ correspond to the \ac{LoS} component and the reflections, respectively. Then, $V_{1,i}$ denotes the square-root of the pathloss, $d_{1,i}$ is the distance from the transmitter to the $i$-th element in the \ac{LIS}; and the term $B$ denotes the scatterers acting as virtual sources where $V_{n,i}$ encapsulates the pathloss and phase shift of all the reflections involved.

%The distance at which we can discern between the far and the near field is given by the Rayleigh distance \cite{zhou2015spherical}:
% \begin{equation}
%     Z = \frac{2D^2}{\lambda}, 
% \end{equation}
% where $D$ is the maximum dimension of the antenna array and $\lambda$ the wavelegth. %\pres{Actually... we are using this distance for literally nothing, right?}

\begin{figure}[t]
    \centering
    \includegraphics[width=0.9\columnwidth]{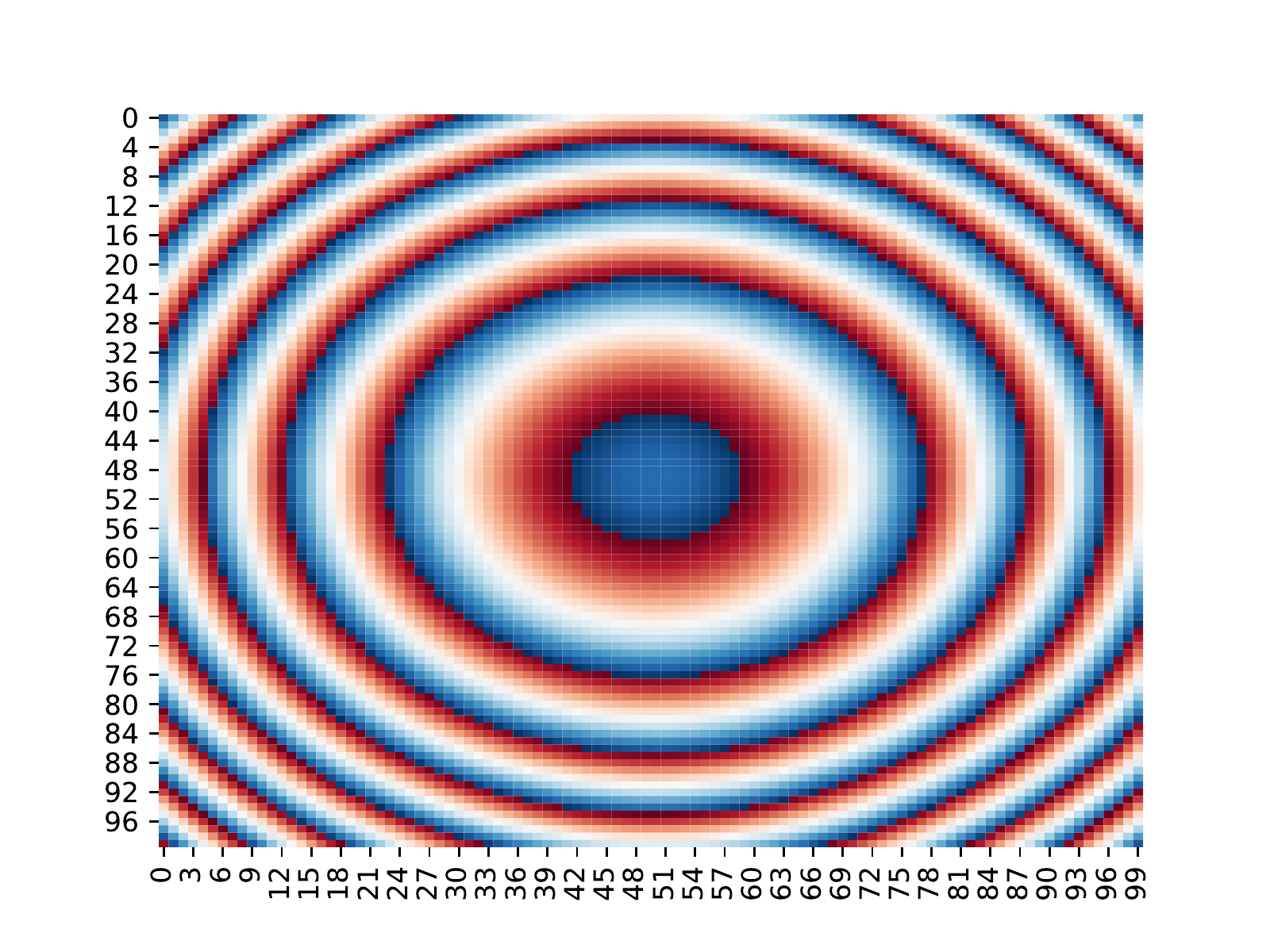}
    \caption{Phase representation of the designed spherical filter based on (\ref{eq:pattern}) for $f=3.5$ GHz, $d=6.2$ m and $N_f=100\times100$ antenna elements $\frac{\lambda}{2}$ spaced \ac{LIS}.}
   \label{fig:MF_pattern}
\end{figure}

\begin{figure}[t]
    \centering
    \includegraphics[width=0.9\columnwidth]{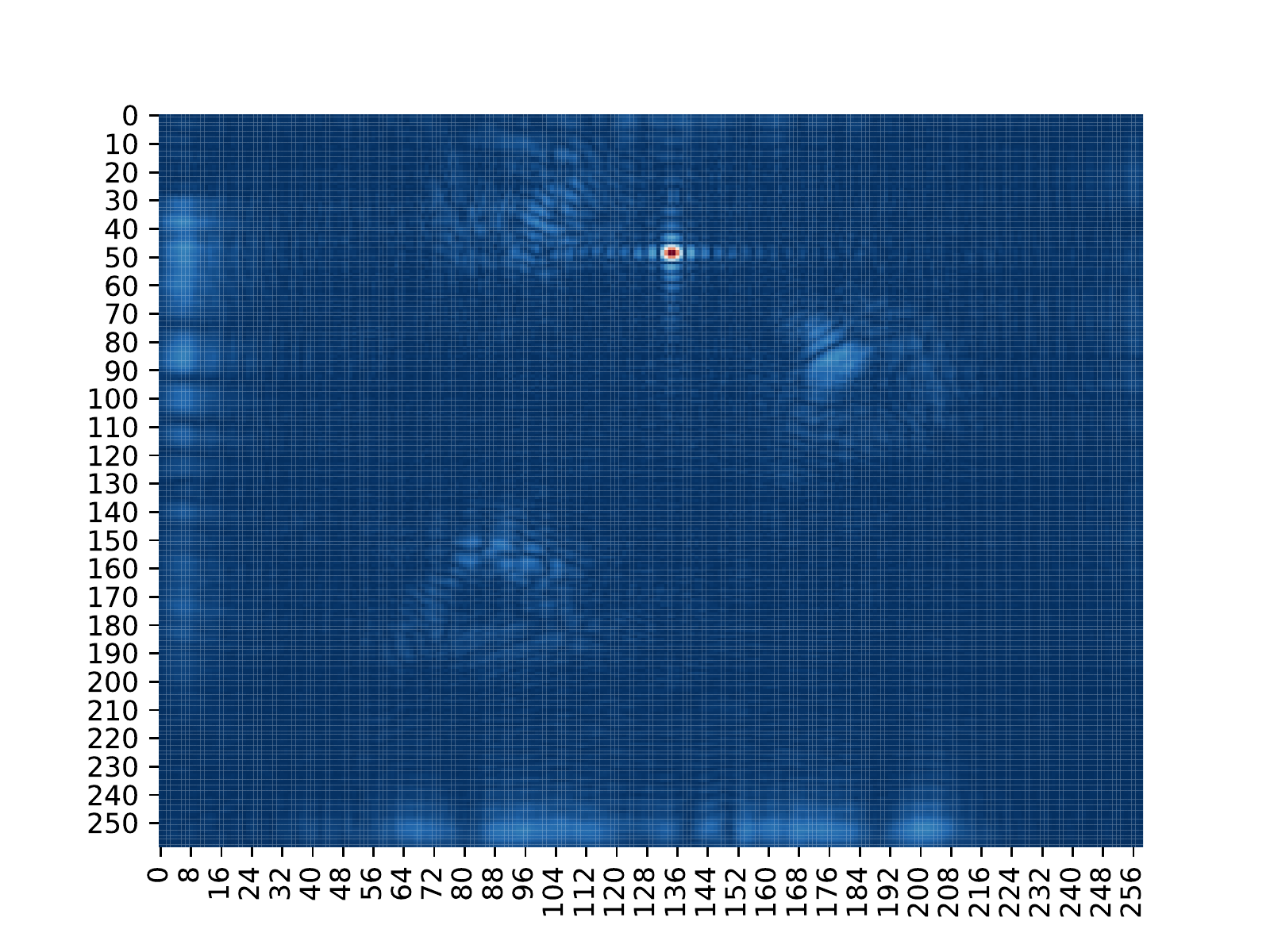}
     \caption{Exemplary radio map obtained for an $M=259\times259$ antenna elements $\frac{\lambda}{2}$ spaced \ac{LIS} in a noiseless scenario with $U_a=1$ users by using the \ac{MF} design represented in Figure \ref{fig:MF_pattern}.}
    \label{fig:MF_map}
\end{figure}

From (2), we can see the spherical-wave channel coefficient $h_{s,i}$ at the \ac{LIS} $i$-th element from an arbitrary user transmission is proportional to \cite{zhou2015spherical} 
\begin{equation} h_{s,i} \ \propto\ \frac{1}{d_i}e^{-j\frac{2\pi}{\lambda}d_i},\label{eq:pattern} \end{equation}
where $d_i$ accounts for the distance from the transmitter to the receiver antenna. %\pres{Mathematically, don't know how to interpret the ratio between an scalar and a vector (element-wise)? Also, I'm wondering whether could be better to explain the channel in terms of sum of paths as usual, where the dependence is clearly observed, i.e.,
%\begin{equation}
%     h_{u,k} = V_{0,k} e^{-j2\pi/\lambda d_{0,k}} + \sum_{n=1}^N V_{n,k} e^{-j2\pi/\lambda d_{n,k}}, 
% \end{equation}
% where $V_{0,k}$ is basically the square-root of the pathloss, $d_{0,k}$ is the distance from the tx to each element in the LIS, and the other terms are the scatterers acting as virtual sources where $V_{n,k}$ encapsulate the pathloss and phase shift of all the reflections but the last one.}
In this way, describing the surface in a vectorized notation, we can derive a \ac{MF} such that:
%\begin{equation} \mathbf{y_f} =  \mathbf{h_s} \ast \mathbf{y}, \end{equation}
\begin{equation} \mathbf{y_{f}}[i] =  \sum_{m=1}^{M}\mathbf{h_s}[i-m]\mathbf{y}[i], \end{equation}
where $i$ denotes the current antenna index, $m$ represents the convolution is done along all the \ac{LIS} dimension. Then, $\mathbf{h_s} \in\mathbb{C}^{N_f\times 1}$ denotes the expected spherical pattern (steering vector) %\pres{I'm not sure about the correctness of the term steering vector here (although I've been using it, sorry about that). All the definitions I've seen about the steering vector assume planar and homogeneous waves, which is not our case} 
for $N_f$ antennas \ac{LIS} deployment on (\ref{eq:pattern}), $\mathbf{y}$ the received signal from (\ref{eq:RecSignal}) and $\mathbf{y_f} \in\mathbb{C}^{M\times 1}$ the filtered output that represents the radio map. As the convolution operator requires both vectors to have the same dimension, we zero-pad $\mathbf{h_s}$ such that it matches $\mathbf{y}$ dimension.  
Fig. \ref{fig:MF_pattern} shows the \textit{pattern} the \ac{MF} operation is using. We can see the expected spherical wave propagation due to the near-field conditions. Please note for the design of the filter, we have to know the frequency $f$ employed and we assume a distance $d$ from the transmitter and the \ac{LIS} receiver in the ceiling. Fig. \ref{fig:MF_map} shows an exemplary radio map obtained by computing $\lvert \mathbf{y_f}\rvert$. In the exemplary scenario, one active transmitter $U_a=1$ is used\footnote{The radio map reconstruction works regardless of the amount of active $U_a$. As denoted in eq (\ref{eq:RecSignal}) they are superposed, meaning the radio map reconstruction can be obtained when $U_a \ge 1$. In this exemplary case, $U_a=1$ was used for illustration purposes.}, while three static scatterers are present in the environment. %These scatterers are modeled as cylinders of metallic materials. 
We see the three scatterers in the environment (the cylindric-like shapes) while we can also identify the highest peak representing the user transmission. The scatterers are captured because from the receiver \ac{LIS} viewpoint, the scatterers act as virtual sources that are equivalent to LoS components, i.e., in \eqref{eq:pattern} the different reflections are equivalent to a LoS path with complex gain $V_{n,i}$. %\pres{we don't correlate, it's implicit from the channel expression as a sum of rays}. 
%Despite of its simplicity, the \ac{LIS} characteristics (dimension and high amount of antennas) allow obtaining a high spatial resolution, leading to a quite accurate description of the environment. %\KK{Last sentence should include the reason why the approach appears effective (spatial selectivity despite simplicity).}

\begin{figure}[t]
    \centering
    \includegraphics[width=0.9\columnwidth]{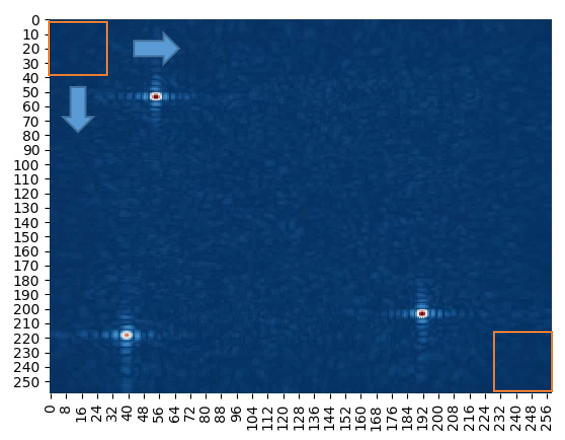}
     \caption{Exemplary radio map obtained for an $M=259\times259$ antenna elements $\frac{\lambda}{2}$ spaced \ac{LIS} in a noisy scenario ($\gamma = 0$ dB) with $U_a=3$ users by using the \ac{MF} design represented in Figure \ref{fig:MF_pattern} and no scatterers presented in the area.}
    \label{fig:MF_sliding}
\end{figure}

\section{Active and Passive multi-user detection based on \ac{LIS} radio map}
\subsection{Detection of multiple active users}
The previous section shows that we can identify the users by locating the local maxima along the entire radio map. In this way, we develop an algorithm to perform that task based on maximum filtering. 
For that, we slide around the image a kernel of size $2 \times K_a + 1$, with a stride of 1, being $K_a$ a parameter we denote as minimum distance (i.e. we assume peaks are separated by at least the minimum distance). In this kernel, we obtain the maximum value of it, considered to be the local maxima of that sub-region. Then, we sort the obtained local maxima in descending order and get the maximas $(x_p, y_p)$ pixel coordinates corresponding to the $U_a$ total number of users. We compare the energy of these peaks to determine how many active regions (peaks) are in the scenario. This is possible because the \ac{MF} output energy in the active users positions is maximized. This means we can detect how many active users are in the map by measuring the energy of the obtained local maximas and compare them sequentially in descending order. When the energy value drops up to 90\%\footnote{This value and $K_a$ were found empirically.}, we know we have detected all the active users. This means we do not need to know how many active users are deployed a priori. Fig.\ref{fig:MF_sliding} shows an example of this method\footnote{For the sake of focusing on the concept of active user detection, we have removed the scatterers in this exemplary map.}. The orange box denotes the kernel of size $2 \times K_a + 1$ that is used to slide along the radio map. Then, the values are sorted in descending order and we infer the 3 users positions. We show a noisy scenario to highlight that because the \ac{MF} maximizes the \ac{SNR} at the output, the users are still nicely captured in the map regardless of a noisy condition.

%\KK{Is the number of active or passive users known at any point? Make this explicit somewhere.}

\subsection{Passive human detection}
%\pres{I would highlight a bit more the importance of detecting people passively, and would sketch the difference opportunities this could bring forward.}

The proposed method assumes that in the scenario, the scatterers are located at fixed positions while the moving entities would correspond to the either active or passive (humans) in the area.

\begin{figure}[t]
\centering
\subfloat[Exemplary clusterized k-means map obtained by using the \ac{MF} map in Figure \ref{fig:MF_map}. \label{fig:clusterized}]{\includegraphics[width=0.23\textwidth]{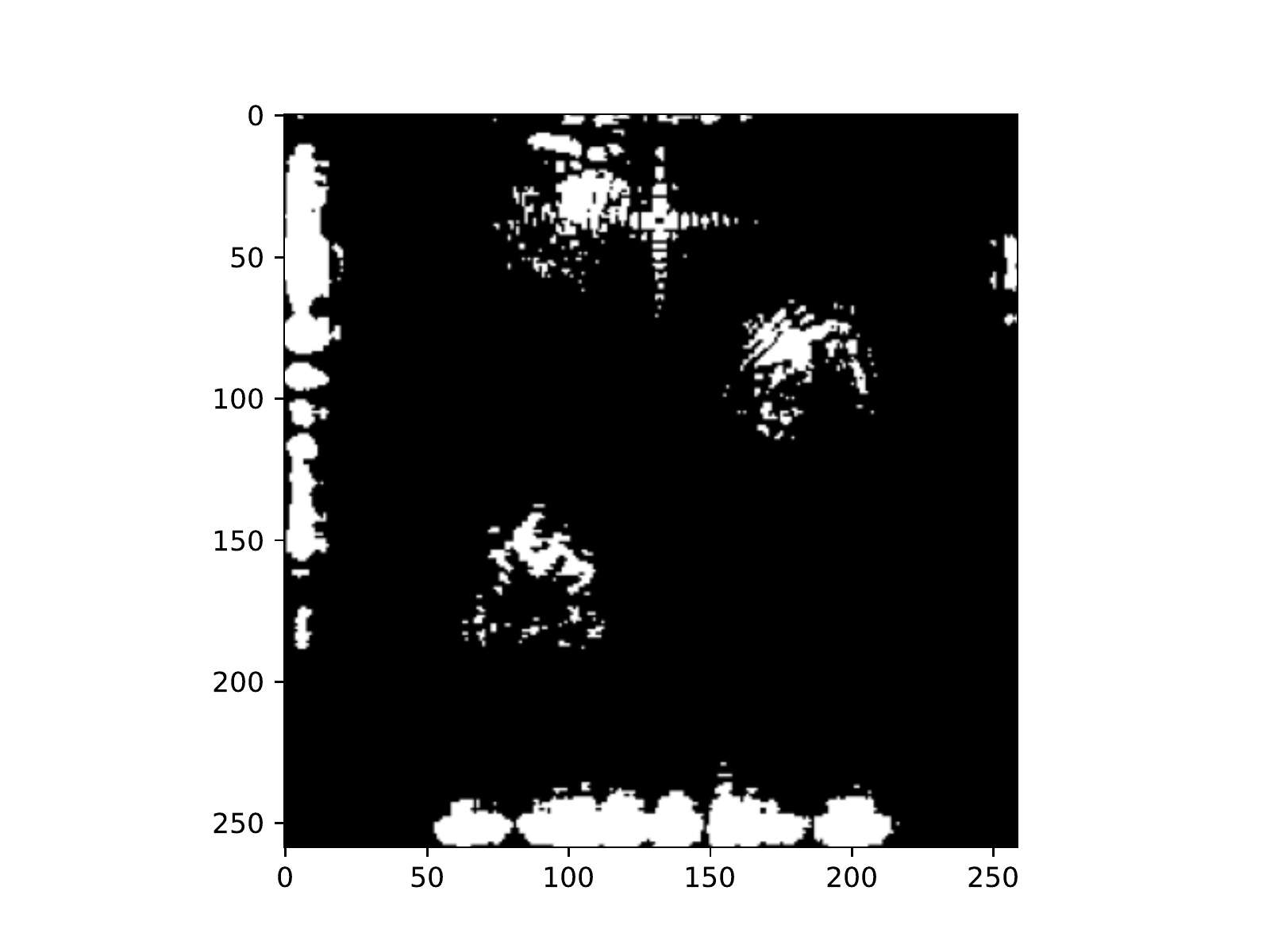}} \hfill
\subfloat[Exemplary clusterized k-means masking map obtained by the combination of $U_a=10$ random transmissions. \label{fig:clusterized_combined}]{\includegraphics[width=0.23\textwidth]{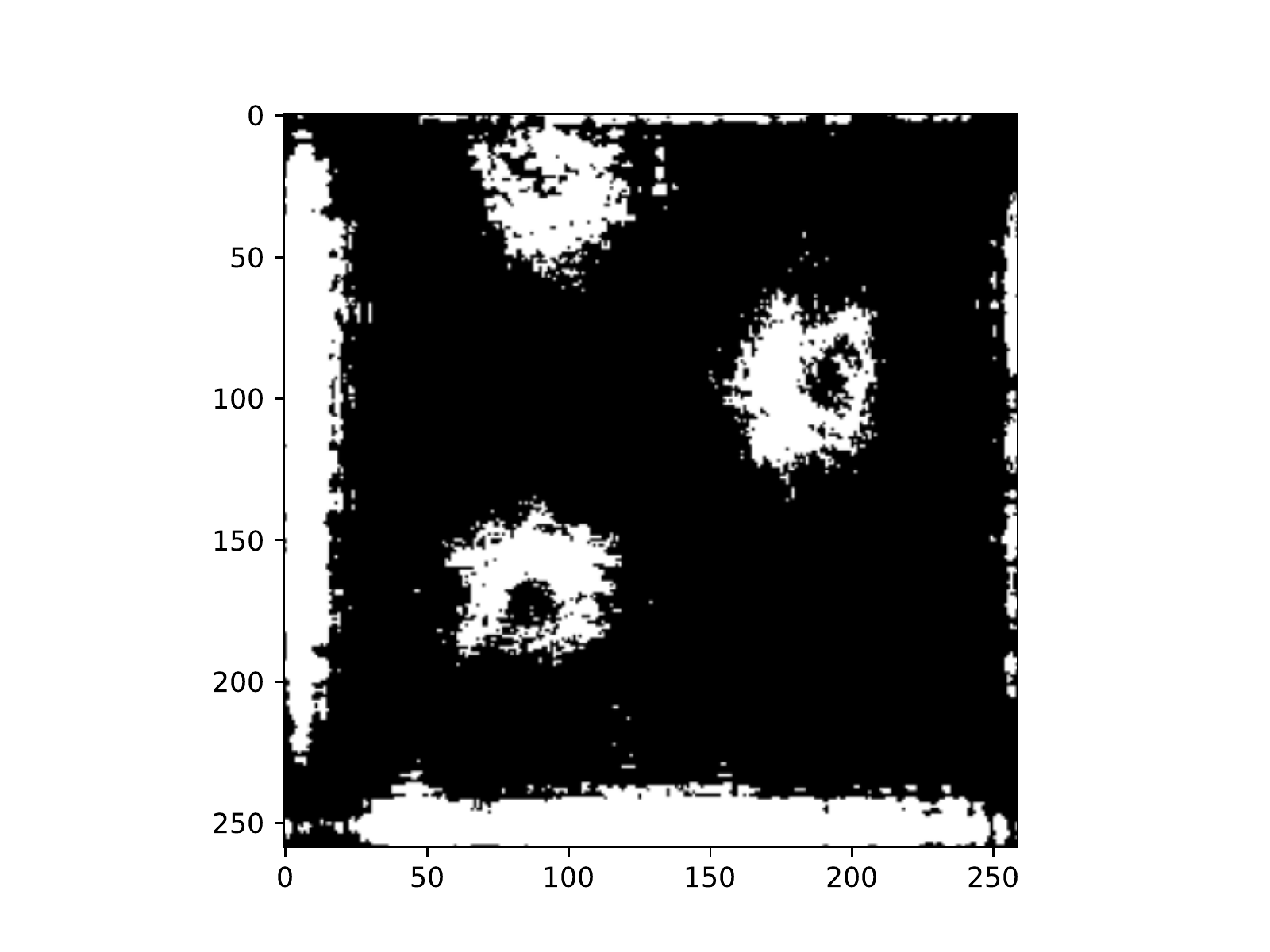}} \hfill
\subfloat[Exemplary clusterized k-means negative masking map obtained by the combination of $U_a=10$ random transmissions. \label{fig:humans_map}]{\includegraphics[width=0.23\textwidth]{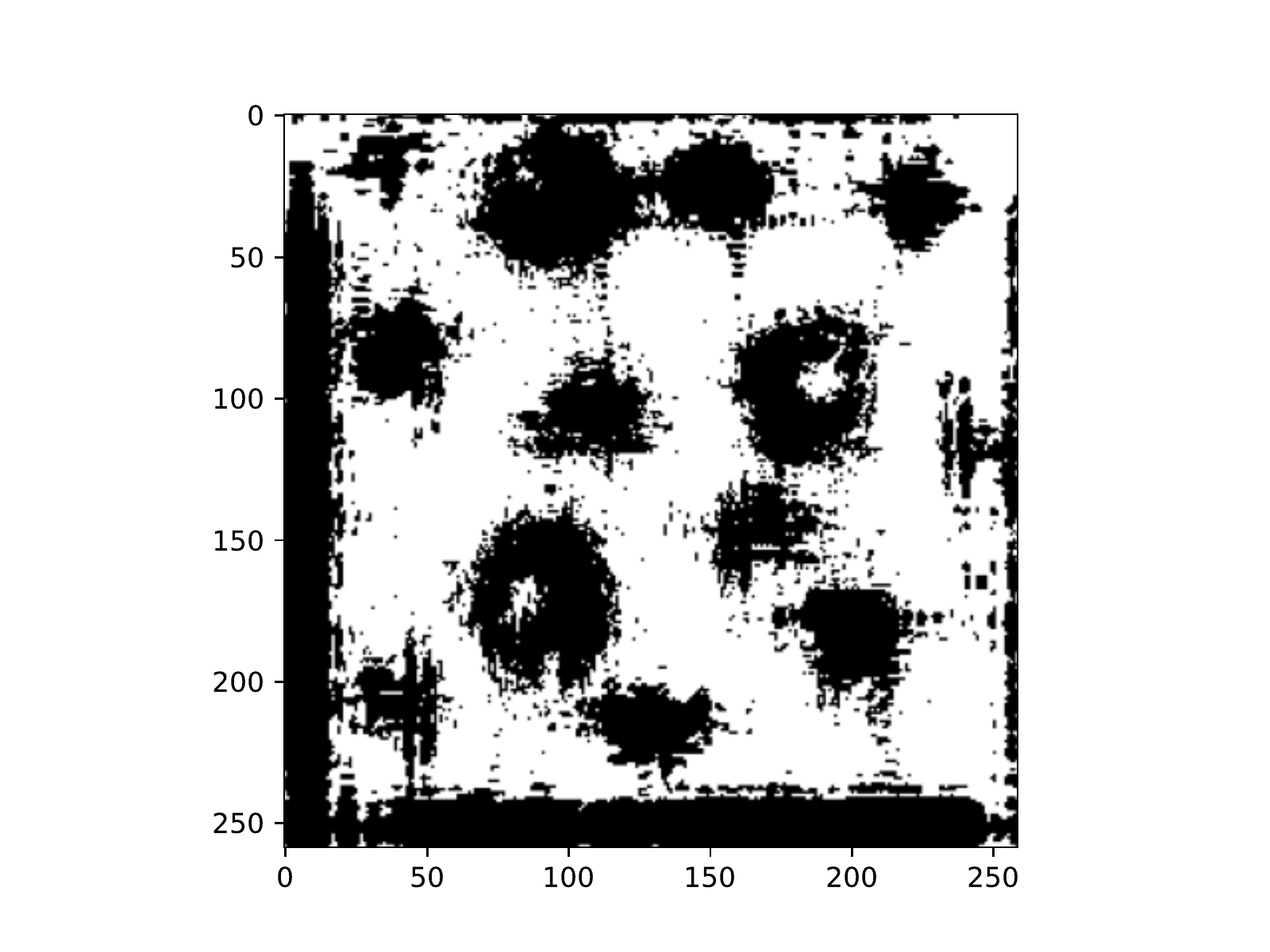}} \hfill
\subfloat[Exemplary logical OR processed negative masking map obtained by the combination of $U_a=10$ random transmissions. \label{fig:humans_map_or}]{\includegraphics[width=0.23\textwidth]{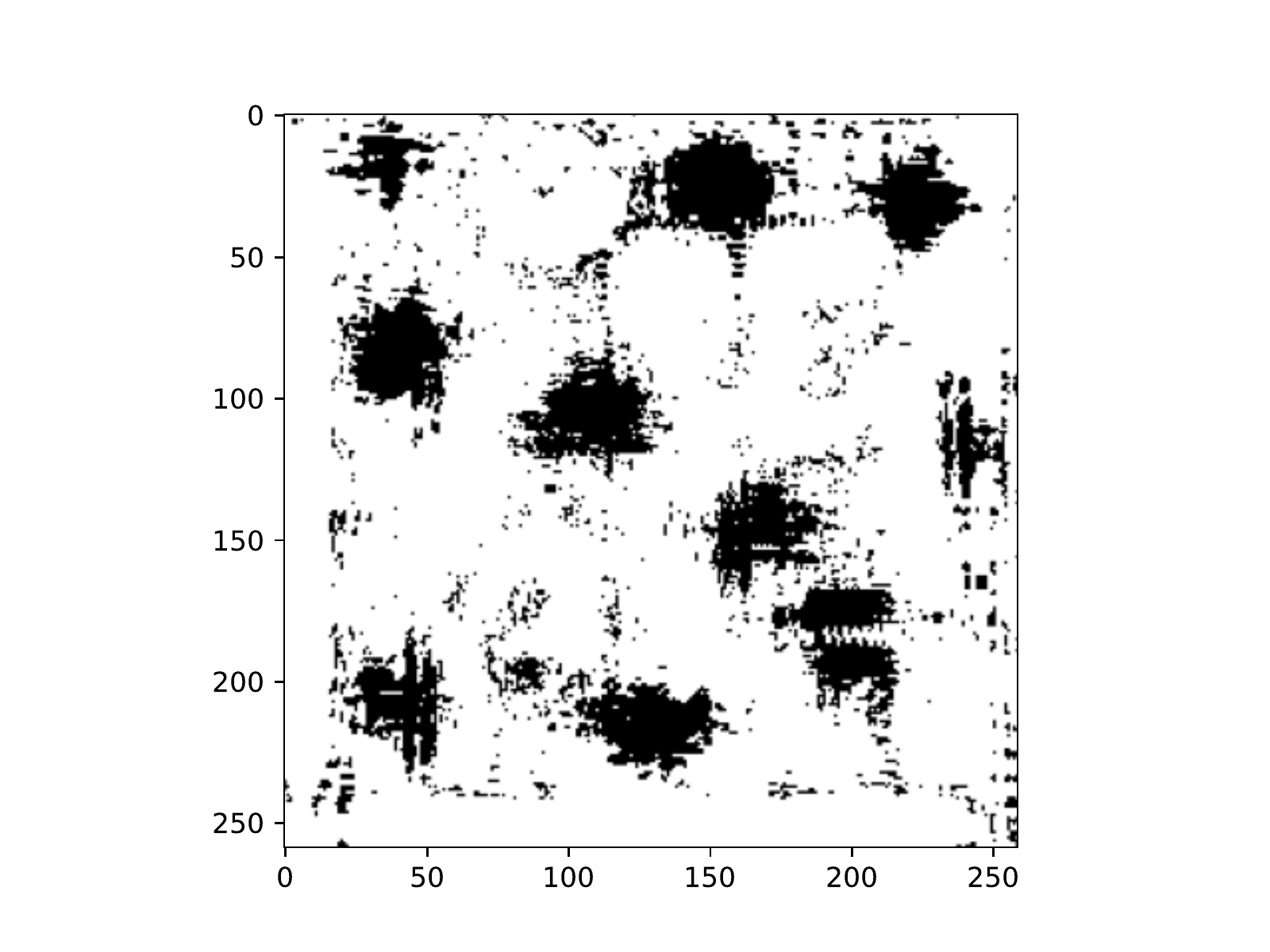}} \hfill
\caption{Radio map processing}
\end{figure}

To perform the passive human detection, we first take advantage of an offline scanning period phase in which we measure different transmissions of any $U_a$ active devices to scan the static features of the propagation environment (a.k.a the scatterers). %\pres{Clearly state this is an offline phase for which you can use any transmitting device.}. 
We then obtain $U_a$ %\pres{$U_a$ refers both to the number of devices and measurements, so we only take one snapshot right?} 
measurements of the environment for different random active user positions when no passive humans are in the scenario. Figure \ref{fig:MF_map} shows that we have mainly two dominant ranges of pixel values, either the background (low energy at the output of the \ac{MF}) or high energy (the active transmitter and  scatterers). This leads us to apply a k-means clustering w.r.t. the pixel values of the radio map (with $k=2$) to enhance the radio map through its binarization. Figure \ref{fig:clusterized} shows a clusterized version of the radio map presented in Figure \ref{fig:MF_map}. It shows the enhanced areas of the static features of the environment as well as the active transmitter. As we are not interested in the active transmitter, we use a technique from the computer vision literature called Template Matching \cite{brunelli2009template} which removes the expected active transmitter pattern from the clusterized map. By combining different active transmissions along the scenario, we can combine several radio maps to obtain an enhanced version that highlights the scatterers presence in the scenario, as shown in Figure \ref{fig:clusterized_combined}. These multiple transmission positions illuminate the scatterers from different angles. Furthermore, these map pixel values are either 0 (black) or 1 (white), being white the representation of the scatterers. We will denote this map as masking map\footnote{A similar strategy could be applied to avoid the necessity of an empty room scanning period, in which the passive humans could be averaged out from the masking map.}. %\KK{Reflect on the assumption of not having humans in the training phase. This is optimistic, but it is relatively simple to have a long enough training (sampling) period such that humans get averaged out in the images. Only fixed objects remain.}

\begin{figure}[t]
\centering
\subfloat[Exemplary groundtruth scenario with $U_p=10$ represented as rectangles. \label{fig:gt_map}]{\includegraphics[width=0.3\textwidth]{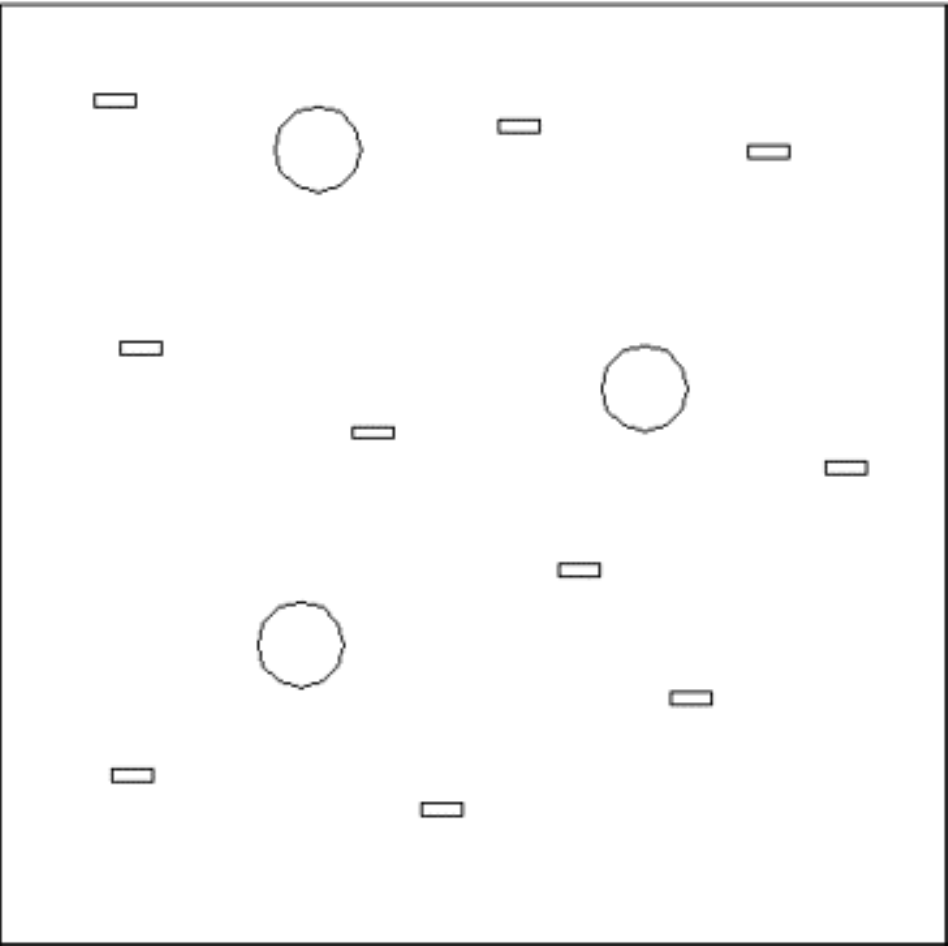}} \hfill
\subfloat[Component Labeling applied to Figure \ref{fig:humans_map_or}. \label{fig:cl}]{\includegraphics[width=0.3\textwidth]{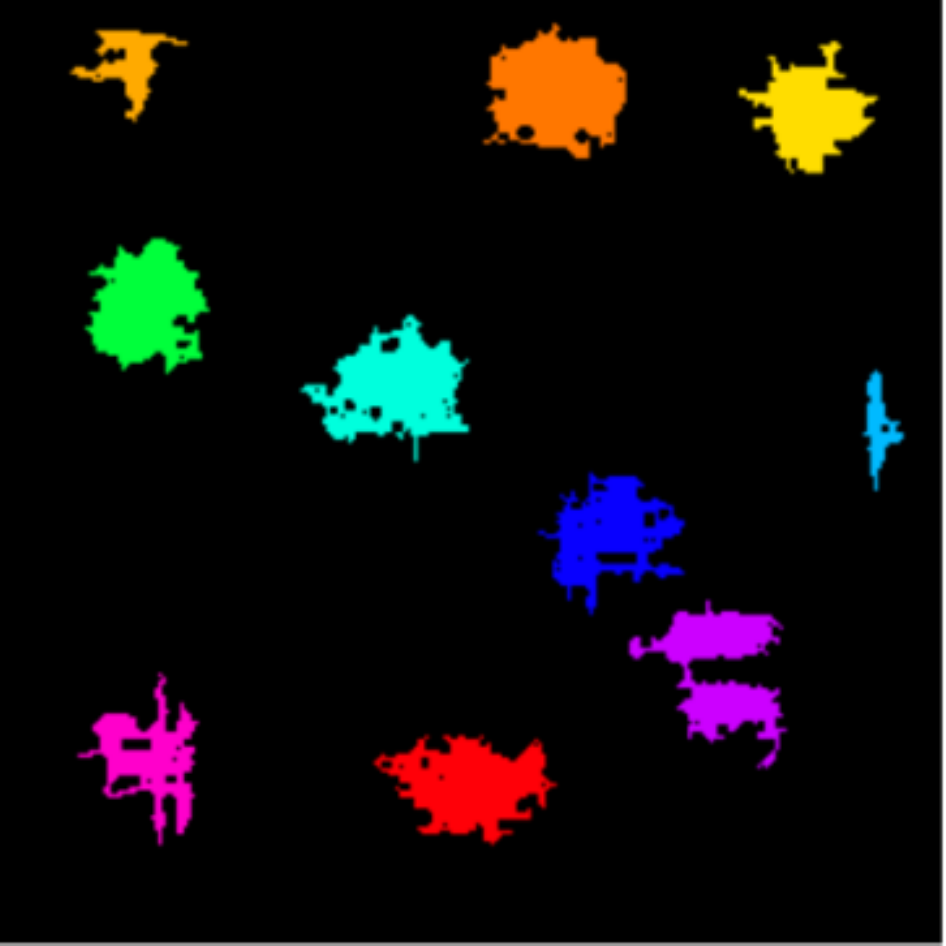}} \hfill
\caption{Groundtruth position of the $U_p$ humans vs the Component Labeling result.}
\end{figure}

By having this representation of the static elements of the environment, we can now store this masking map locally at the \ac{LIS} to process new maps and remove the static elements of it when trying to detect humans passively. For this purpose, when there are passive humans in the room, we can obtain combined radio maps to process a negative masking map (meaning scatterers are now black) that we will use to perform a logical OR operation with the locally stored masking map. Figure \ref{fig:humans_map} shows an example of a negative masking map when there is $U_p=10$ passive humans in the scenario. We see now the scatterers and the humans are represented in black (0 value). Furthermore, when applying an OR operation with the locally stored masking map we obtain Figure \ref{fig:humans_map_or}, which eliminates the static scatterers of the scenario and highlights the passive humans reflections. We can see in the map that there are some artifacts (salt-pepper noise) as a result of this process. To alleviate it, we define a sliding window algorithm of size $K_c \times K_c$ that set all the pixel values comprising the window size to 1 (white) if the number of black pixels in that window is lower than a defined threshold $T_h$. In this way, we can reduce significantly this salt-pepper noise. Figure \ref{fig:cl} shows the removal of the artifacts thanks to this procedure.

Finally, we are interested in detecting these shapes associated to the passive human positions in the radio maps. For that, we adopt a computer vision algorithm named Component Labeling \cite{rosenfeld1966sequential} which compares neighboring pixels to detect a shape that is assigned to the same label. Figure \ref{fig:gt_map} shows the exemplary groundtruth scenario in which these maps are computed while Figure \ref{fig:cl} shows the result of detecting the $U_p=10$ passive humans. They are assigned to different colors (labels) for illustration purposes. Hence, we can infer the passive human positions by obtaining the center pixel coordinates of these shapes $(x_p, y_p)$.  %\KK{Repeat: treat possible uses cases separately (intro).} %\pres{I like this explanation =) I miss also pointing out a little bit how these maps could be used (not only the detected users, but also the scatterers) apart from the passive human application we present here. I think we need to highlight that the idea is more powerful and not limited to what we present here.}

\subsection{Position mapping from pixel to real space}
In both presented methods, we infer the positions in the pixel domain. To infer the real position of the users, we have to perform a mapping described as
\begin{equation}
    c = c_p \times \Delta s,
\end{equation}
where $c \in \{x, y\}$ is the real inferred mapped coordinate,  $c_p \in \{x_p, y_p\}$ refers to the inferred pixel coordinate and $\Delta s$ corresponds to the inter-antenna spacing.

\section{Exemplary scenario}
\subsection{Simulated scenario}
\label{sec:Scenario}
 We conducted simulations via ray tracing \cite{FEKO}. The motivation for using ray tracing is twofold:
 \begin{enumerate}
     \item Providing a channel that is not generated from the same model we used to analyze the system. In this way, we can guarantee a match between expected theoretical results and empirical ones.
     \item We can control the geometry of the environment to test different conditions.
 \end{enumerate}
  We simulate a scenario of size $10.34 \times 10.34 \times 8$ m. We deploy an \ac{LIS} with $259\times259$ elements separated $\lambda/2$. Each $U_a$ active device transmits an arbitrary narrowband signal of 20 dBm at 3.5 GHz. The distance from which the \ac{MF} is calibrated is $d=6.2$. The active $U_a$ are assumed to be $\ge 1.8$ m height, being this value randomnly selected. The scatterers are modeled as metallic (with conductivity $s=19444$ S/m, relative permittivity $\epsilon=1$ and relative permeability $\mu=20)$\footnote{These values are provided by the software manual \cite{FEKO}.} cylinders of 1 m diameter and 2 m height. The passive $U_p$ humans are model as rectangles of dimensions 0.3x0.5x1.7 m (average human dimensions obtained from \cite{potkany2018requirements}) with $s=1.44$ S/m, $\epsilon=38.1$ and $\mu=1$ \cite{hall2007antennas}.%\pres{I would describe these parameters just in a paragraph. E.g., We consider an scenario of dimensions $10.34x10.34x8 m$, with a LIS of 259x259 elements separated $\lambda/2$. Each active device transmits an arbitrary narrowband signal of 20 dBm at 3.5 GHz... }

% \begin{table}[h!]
% \caption{Parameters}
% \label{table_parameters}
% \resizebox{\columnwidth}{!}{
% \begin{tabular}{|c|c|c|c|c|c|}
% \hline
% \begin{tabular}[c]{@{}c@{}}Frequency \\ (GHz)\end{tabular} & \begin{tabular}[c]{@{}c@{}}Tx \\ Power \\ (dBm)\end{tabular} & \begin{tabular}[c]{@{}c@{}}Nray \\ paths\end{tabular} & \begin{tabular}[c]{@{}c@{}}Antenna \\ type\end{tabular} & \begin{tabular}[c]{@{}c@{}}Surface dimension \\ (LxWxH m)\end{tabular} & \begin{tabular}[c]{@{}c@{}}Antenna spacing \\ (cm) \end{tabular} \\ \hline
% 3.5                                                        & 20                                                           & All significant paths                                                    & Omni                                                    & 10.34x10.34x8                                                                                                      & $\frac{\lambda}{2}$                                                              \\ \hline
% \end{tabular}}
% \end{table}
 %The most relevant parameters used for simulation are summarized in Table \ref{table_parameters}. \pres{Can't we put the parameters in another way? It's hard to read the table}

\subsection{Received signal and noise modeling}
From the ray-tracing simulation, the received signal in (\ref{eq:RecSignal}) is obtained as the complex electric field arriving at the $i$-th antenna element, $\widetilde{E}_{i}$, which can be regarded as the superposition of each ray path from every $u \in U_a$ user, i.e., 
\begin{equation}
    \label{eq:Esum}
  \widetilde{E}_{i} = \sum_{u=1}^{U_a}\sum_{n=1}^{N_r} \widetilde{E}_{i,n,u}= \sum_{u=1}^{U_a}\sum_{n=1}^{N_r}E_{i,n,u} e^{j\phi_{i,n,u}}.  
\end{equation}
Then, the complex signal at the output of the $i$-th element is therefore given by %\pres{Relate this to (1) please, specially to the channel}
\begin{equation}
    \label{eq:complexSignal}
    y_i = \sqrt{\frac{\lambda^2Z_i}{4\pi Z_0}} \widetilde{E}_{i} + n_i,
\end{equation}
with $\lambda$ the wavelength, $Z_0 = 120\pi$ the free space impedance and  $Z_i$ the antenna impedance. %\pres{Use same subindex and notation than in (1)} %and $n_i$ is complex Gaussian noise with zero mean and variance $\sigma^2$ \pres{This has been already explained}. %Note that \eqref{eq:complexSignal} is exactly the same model than \eqref{eq:RecSignal}. 
For simplicity, we consider $Z_i = 1\,\forall\, i$.  Finally, in order to test the system performance under distinct noise conditions, the average \ac{SNR}, $\gamma$, is defined as
\begin{equation}
    \label{eq:snr}
    \gamma \triangleq  \frac{\lambda^2}{4\pi Z_0 M \sigma^2}\displaystyle\sum_{i=1}^{M} |\widetilde{E}_{i}|^2,
\end{equation}
where $M$ denotes the number of antenna elements in the \ac{LIS}.

%\subsection{Noise averaging strategy}

%The presence of noise may be critical in the radio image sensing, since it impacts considerably in the image classification performance \cite{roy2018effects}. 

%Trying to mitigate the noise impact, let us assume the system can obtain $S$ extra samples at each channel coherence interval to perform an S-averaging. Note that, if $S\to\infty$, then 
%\begin{equation}
%  \left. y'_{i}\right|_{S\rightarrow\infty} = \mathbb{E}[y_{i} | h_{i}] = \sum_{u=1}^{U_a}h_{u,i},
%\end{equation}
%meaning that the noise variance at the resulting image has vanished, i.e., the received superimposed signal at each antenna (conditioned on the channel) is no longer a random variable. Observe that in this way the  image  preserves  the  pattern. This effect is only possible if the system  is able  to  obtain  a  very  large  number $S$ of samples within each channel coherence interval. %\pres{Again, I'm wondering if this is necessary, or can be just summarized saying that we do averaging resulting in an improvement of the average SNR by a factor of $S$. } %\KK{I agree. The value of S here is actually the legth of the pilot sequence used for sensing.}

\section{Numerical results and Discussion}
%For the following results, we deploy an \ac{LIS} in the ceiling whose physical aperture is the size of the scenario, comprised by a $M=259 \times 259$ antennas array. %\pres{That's the number of antennas, not the physical aperture. Clearly state that the physical aperture is exactly the size of the scenario.} 
We evaluate the performance of the proposed method in the scenario described in Section \ref{sec:Scenario} under different $\gamma$ conditions. We assume the system can obtain $S$ extra samples at each channel coherence interval to perform an S-averaging, diminishing the noise variance contribution.
\subsection{Active user detection}
We here present as a baseline, a comparison of the sensing performance by tracking $U_a=3$ users simultaneously transmitting under different $\gamma = 10/0$ dB scenarios. For these results, we set the minimum distance of the sliding window to $K_a=5$. 

\begin{figure}[t]
    \centering
    \includegraphics[width=0.95\columnwidth]{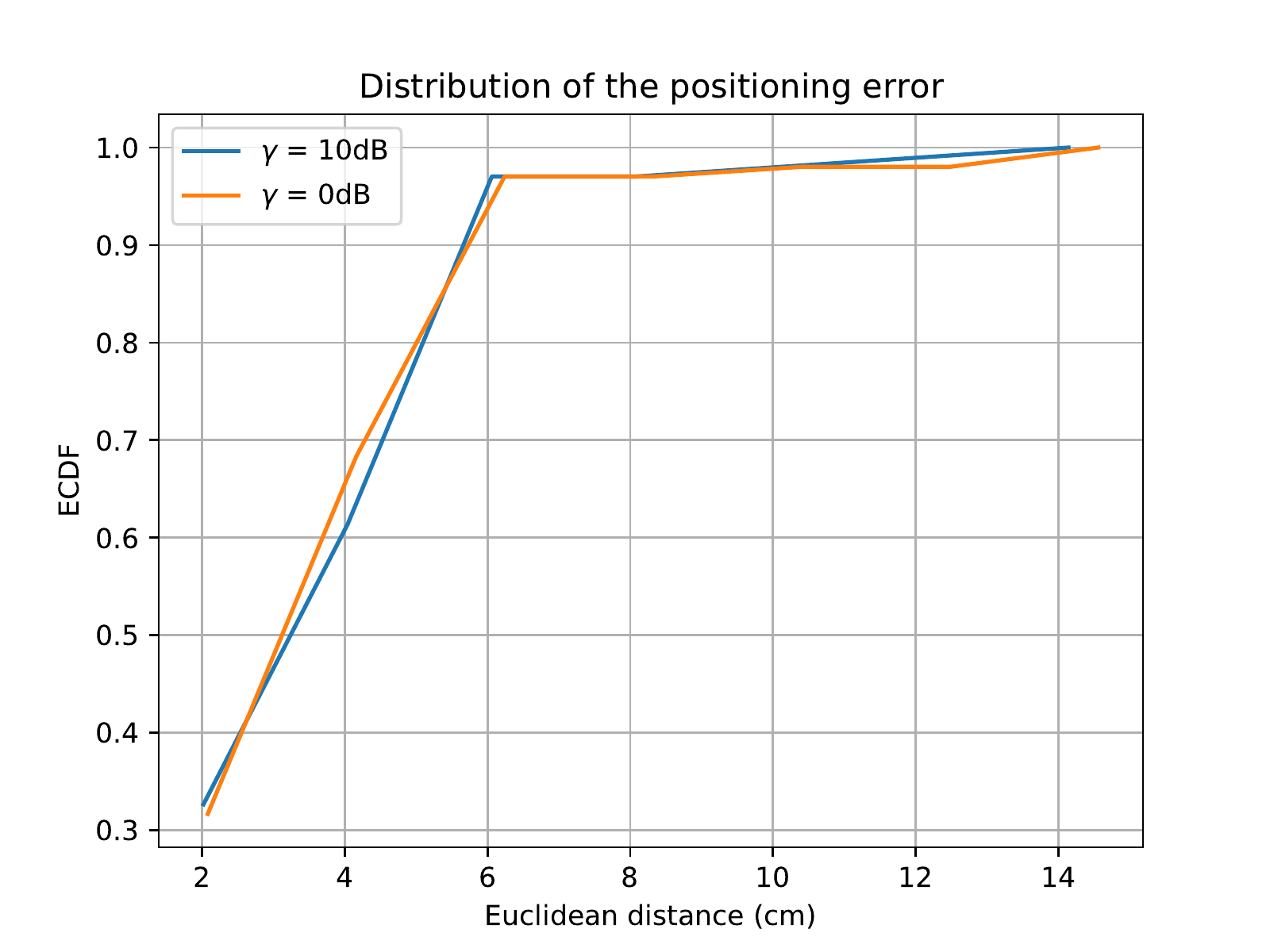}
    \caption{ECDF for radio image sensing for $U_a=3$ active transmissions, with fixed \ac{LIS} aperture of $M=259\times 259$.}\label{fig:active}
\end{figure}

Figure \ref{fig:active} shows the \ac{ECDF} of the euclidean distance between the predicted and the groundtruth positions for a $U_a=3$ simultaneous transmission. The results show the sensing system can perform a multi-user detection with a quite high accuracy in positioning. This is thanks to the \ac{LIS} physical aperture, which allows obtaining a high resolution image that creates a pattern discernable while 3 robots are transmitting in the scenario. We can see the positioning error is around 5.5 cm under 90\% of the cases, which shows its tracking potential. Also, as we are relying on the \ac{MF} output which maximizes the \ac{SNR}, the robustness of the system to noisy scenarios is significant, leading to similar results for both the $\gamma = 10$ and $\gamma = 0$ dB cases. %\pres{As discussed, maybe this result can be deleted and a more insightful comparison can be introduced.}

\subsection{Passive user detection}
We here leverage the performance for passive user detection in the scenario using the method described in Section IV.B. We fix $\gamma=0$ dB. For these results, we set $K_c=2$ for the sliding window algorithm, $T_h=0.5$, $S=100$ averaged measurements and we consider $U_p = 10$ humans at arbitrary positions in the scenario. %\pres{And consider 10 humans at arbitrary positions? I sounds weird to me "to deploy a human"}.

\begin{figure}[t]
    \centering
    \includegraphics[width=0.95\columnwidth]{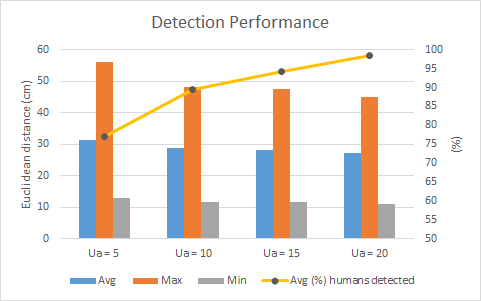}
    \caption{Average human detection percentage (\%) and positioning errors (cm) with fixed \ac{LIS} aperture of $M=259\times 259$, in a $\gamma=0$ dB condition, with $S=100$ averaging strategy and $U_p=10$ humans in the scenario.}\label{fig:passive}
\end{figure}

\begin{figure}[t]
    \centering
    \includegraphics[width=0.95\columnwidth]{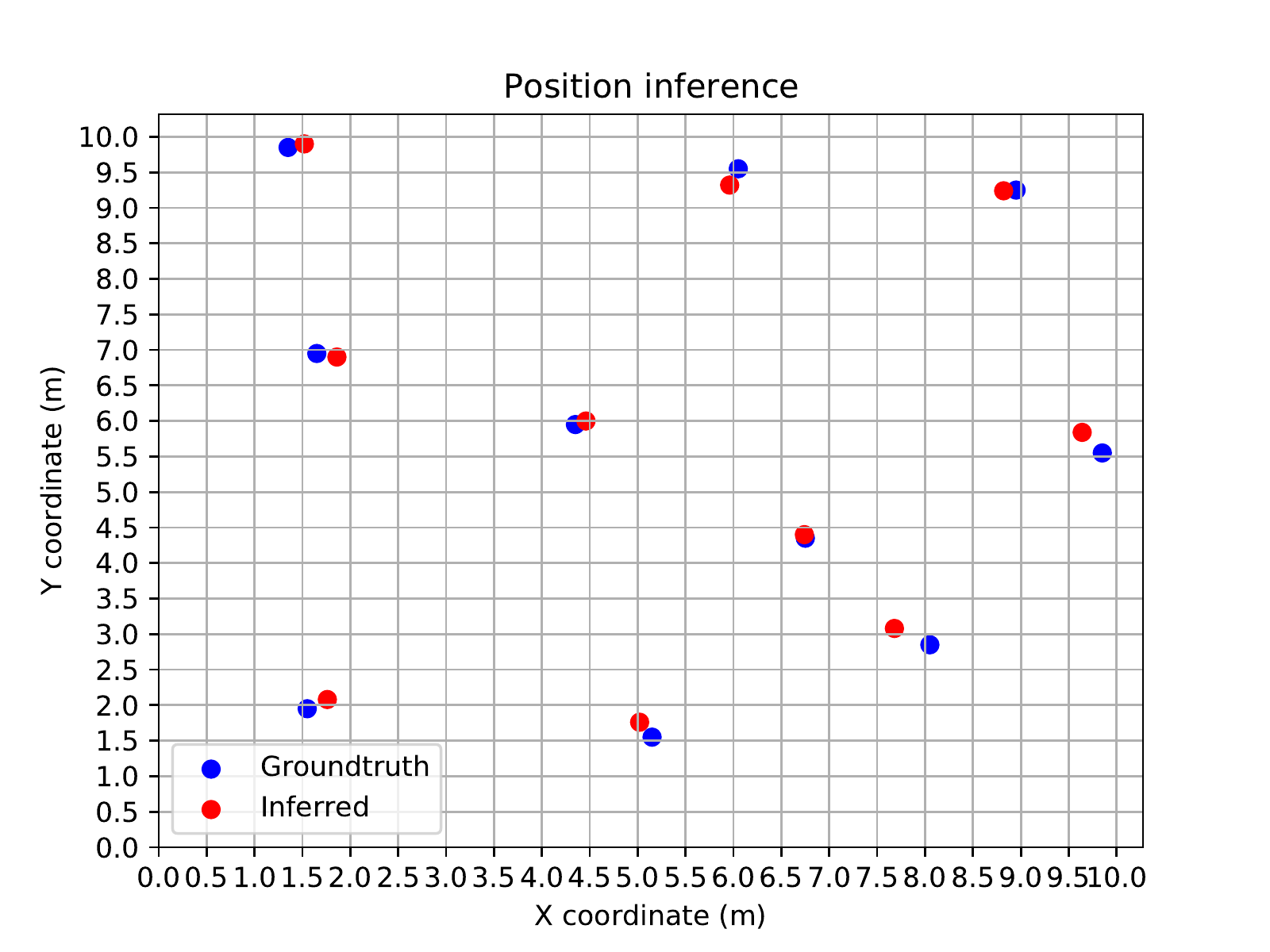}
    \caption{Exemplary human detection with fixed \ac{LIS} aperture of $M=259\times 259$, in a $\gamma=0$ dB condition, with $S=100$ averaging strategy, $U_a=20$ active users and $U_p = 10$ humans in the scenario.}\label{fig:distance_montecarlo_ex}
\end{figure}

The detection of passive humans is highly impacted by the $U_a$ active users positions. For the sake of generalization, we perform Monte Carlo simulations for obtaining our results under different random configurations. Figure \ref{fig:passive} shows the average, maximum and minimum positioning errors of the correctly detected passive humans as well as the average detected humans by using a different number of active users $U_a$. Please note, we are not using dedicated active transmissions for this task, but we take advantage of the wireless communications occurring from these active devices in the scenario. The results show that the number of active users does not really impact on the positioning performance as it remains similar when using a lower and a higher number of active users $U_a$. However, by increasing $U_a$, the number of passive humans detected increases. This is because the more the transmissions, the more reflections we obtain from the human body reflections leading to an easier detection of the passive humans. Furthermore, the detection of this system is quite accurate, as we can detect a minimum of around 80\% humans in all the configurations and the average error is around 28 cm. Figure \ref{fig:distance_montecarlo_ex} shows an illustration of the inferred positions w.r.t. the groundtruth positions. It shows the positioning accuracy is quite high even with 10 people passively sensed. We would like to remark that it is not necessary having simultaneous transmission from all the active $U_a$ users and the performance would remain similar if we wish to combine signals within a defined coherence time. However, we consider the presented case is more challenging (due to interference) and we wanted to show it still works with no drops in performance.  %\pres{One interesting thing here is... do we need all the transmissions to be at the same time? or can we combine the different snapshots using the methods described in previous sections? Because this will allow to exploit the different devices transmitting at different time in a given temporal window (for which the scenario is assumed static)}
\subsection{Passive user detection distance evaluation}
Finally, we here evaluate the accuracy of the detection of passive humans by comparing performance under different separations among them. We fix $\gamma=0$ dB. For these results, we set $K_c=2$ for the sliding window algorithm, $T_h=0.5$, $S=100$ averaged measurements and we consider $U_p = 2$ humans separated 25/50/75/100 cm apart, respectively. 

\begin{figure}[t]
    \centering
    \includegraphics[width=0.95\columnwidth]{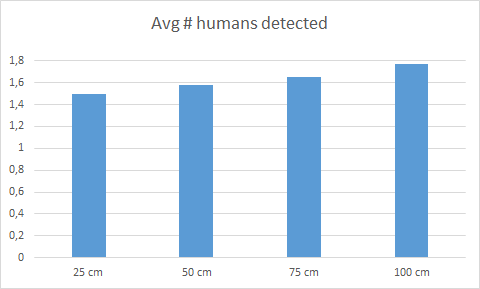}
    \caption{Average human detection with fixed \ac{LIS} aperture of $M=259\times 259$, in a $\gamma=0$ dB condition, with $S=100$ averaging strategy, $U_a=20$ active users and $U_p = 2$ humans in the scenario.}\label{fig:distance_montecarlo}
\end{figure}

We test different separations and we evaluate the detection performance of the $U_p=2$ passive humans. Figure \ref{fig:distance_montecarlo} shows the average detection of the humans.
We can see the system achieves around 1.5/2 detections in the most challenging case (25 cm) while obtaining around 1.8/2 in the most favorable (100 cm). This shows the potential of the system, even when the separation among humans is quite small.

\section{Conclusions}
Large Intelligent Surfaces are a key ingredient in current studies for improving communications in the forthcoming 6G paradigm. However, one of the main characteristics of 6G resides in the ability of sensing. %LIS are a really useful tool for capturing the radio propagation environment as we have demonstrated this technology allows to represent the sense information as an image, something that decreases complexity and permits to use image processing tools for studying the characteristics of the radio environment. What is more, 
The presented use case shows computer vision algorithm along with clustering-based machine learning, are a powerful tool to take into account when using an image-based \ac{LIS} sensing approach. Future lines can be of interest, for example, location assisted analog beamforming according to the user position radio map. %Besides, %a further analysis of the images can be of interest for determining the optimal pattern of antenna elements deployment along the whole surface. 
%autoencoders for image super resolution \cite{vaca2021assessing} can be of interest for reducing the physical aperture of the \ac{LIS} deployed in the ceiling. 
Finally,  we note that \ac{LIS} is one of the technologies being considered for future 6G systems, which may change the relevant cost/benefit analysis in that any sensing functionality is then expected to be added onto the system rather than requiring explicit investment on extra dedicated hardware.

%\newpage

\bibliographystyle{./IEEEtran}
\bibliography{./biblio}   

\begin{thebibliography}{10}
\providecommand{\url}[1]{#1}
\csname url@samestyle\endcsname
\providecommand{\newblock}{\relax}
\providecommand{\bibinfo}[2]{#2}
\providecommand{\BIBentrySTDinterwordspacing}{\spaceskip=0pt\relax}
\providecommand{\BIBentryALTinterwordstretchfactor}{4}
\providecommand{\BIBentryALTinterwordspacing}{\spaceskip=\fontdimen2\font plus
\BIBentryALTinterwordstretchfactor\fontdimen3\font minus
  \fontdimen4\font\relax}
\providecommand{\BIBforeignlanguage}[2]{{%
\expandafter\ifx\csname l@#1\endcsname\relax
\typeout{** WARNING: IEEEtran.bst: No hyphenation pattern has been}%
\typeout{** loaded for the language `#1'. Using the pattern for}%
\typeout{** the default language instead.}%
\else
\language=\csname l@#1\endcsname
\fi
#2}}
\providecommand{\BIBdecl}{\relax}
\BIBdecl

\bibitem{zhang2017indoor}
Y.~Zhang, L.~Deng, and Z.~Yang, ``Indoor positioning based on {FM} radio
  signals strength,'' in \emph{2017 First International Conference on
  Electronics Instrumentation \& Information Systems (EIIS)}.\hskip 1em plus
  0.5em minus 0.4em\relax IEEE, 2017, pp. 1--5.

\bibitem{chiou2009design}
Y.-S. Chiou, C.-L. Wang, S.-C. Yeh, and M.-Y. Su, ``Design of an adaptive
  positioning system based on {WiFi} radio signals,'' \emph{Computer
  Communications}, vol.~32, no. 7-10, pp. 1245--1254, 2009.

\bibitem{song2019efficient}
X.~Song, S.~Haghighatshoar, and G.~Caire, ``Efficient beam alignment for
  millimeter wave single-carrier systems with hybrid {MIMO} transceivers,''
  \emph{IEEE Transactions on Wireless Communications}, vol.~18, no.~3, pp.
  1518--1533, 2019.

\bibitem{noh2017multi}
S.~Noh, M.~D. Zoltowski, and D.~J. Love, ``Multi-resolution codebook and
  adaptive beamforming sequence design for millimeter wave beam alignment,''
  \emph{IEEE Transactions on Wireless Communications}, vol.~16, no.~9, pp.
  5689--5701, 2017.

\bibitem{latva2019key}
M.~Latva-aho and K.~Lepp{\"a}nen, ``Key drivers and research challenges for
  {6G} ubiquitous wireless intelligence (white paper),'' \emph{6G Flagship
  research program, University of Oulu, Finland}, 2019.

\bibitem{adib2014real}
F.~Adib, Z.~Kabelac, H.~Mao, D.~Katabi, and R.~C. Miller, ``Real-time breath
  monitoring using wireless signals,'' in \emph{Proceedings of the 20th annual
  international conference on Mobile computing and networking}, 2014, pp.
  261--262.

\bibitem{stasiak2017fmcw}
K.~Stasiak and P.~Samczynski, ``{FMCW} radar implemented in {SDR} architecture
  using a {USRP} device,'' in \emph{2017 Signal Processing Symposium
  (SPSympo)}.\hskip 1em plus 0.5em minus 0.4em\relax IEEE, 2017, pp. 1--5.

\bibitem{wilson2010radio}
J.~Wilson and N.~Patwari, ``Radio tomographic imaging with wireless networks,''
  \emph{IEEE Transactions on Mobile Computing}, vol.~9, no.~5, pp. 621--632,
  2010.

\bibitem{lee2019variational}
D.~Lee and G.~B. Giannakis, ``A variational bayes approach to adaptive radio
  tomography,'' \emph{arXiv preprint arXiv:1909.03892}, 2019.

\bibitem{wang2016rt}
H.~Wang, D.~Zhang, Y.~Wang, J.~Ma, Y.~Wang, and S.~Li, ``{RT-Fall}: A real-time
  and contactless fall detection system with commodity {WiFi} devices,''
  \emph{IEEE Transactions on Mobile Computing}, vol.~16, no.~2, pp. 511--526,
  2016.

\bibitem{pu2013whole}
Q.~Pu, S.~Gupta, S.~Gollakota, and S.~Patel, ``Whole-home gesture recognition
  using wireless signals,'' in \emph{Proceedings of the 19th annual
  international conference on Mobile computing \& networking}, 2013, pp.
  27--38.

\bibitem{larsson2014massive}
E.~G. Larsson, O.~Edfors, F.~Tufvesson, and T.~L. Marzetta, ``Massive {MIMO}
  for next generation wireless systems,'' \emph{IEEE communications magazine},
  vol.~52, no.~2, pp. 186--195, 2014.

\bibitem{vaca2021assessing}
C.~J. Vaca-Rubio, P.~Ramirez-Espinosa, K.~Kansanen, Z.-H. Tan, E.~De~Carvalho,
  and P.~Popovski, ``Assessing wireless sensing potential with large
  intelligent surfaces,'' \emph{IEEE Open Journal of the Communications
  Society}, vol.~2, pp. 934--947, 2021.

\bibitem{su2001modeling}
T.~Su and H.~Ling, ``On modeling mutual coupling in antenna arrays using the
  coupling matrix,'' \emph{Microwave and Optical Technology Letters}, vol.~28,
  no.~4, pp. 231--237, 2001.

\bibitem{zhou2015spherical}
Z.~Zhou, X.~Gao, J.~Fang, and Z.~Chen, ``Spherical wave channel and analysis
  for large linear array in los conditions,'' in \emph{2015 IEEE Globecom
  Workshops (GC Wkshps)}.\hskip 1em plus 0.5em minus 0.4em\relax IEEE, 2015,
  pp. 1--6.

\bibitem{brunelli2009template}
R.~Brunelli, \emph{Template matching techniques in computer vision: theory and
  practice}.\hskip 1em plus 0.5em minus 0.4em\relax John Wiley \& Sons, 2009.

\bibitem{rosenfeld1966sequential}
A.~Rosenfeld and J.~L. Pfaltz, ``Sequential operations in digital picture
  processing,'' \emph{Journal of the ACM (JACM)}, vol.~13, no.~4, pp. 471--494,
  1966.

\bibitem{FEKO}
Feko, altair engineering, inc. \url{ https://www.altairhyperworks.com/feko}.

\bibitem{potkany2018requirements}
P.~Potk{\'a}ny, M.~Debn{\'a}r, M.~Hitka, and M.~Gejdo{\v{s}}, ``Requirements
  for the internal layout of wooden house from the point of view of ergonomics
  changes,'' \emph{Zeszyty Naukowe. Quality. Production. Improvement}, 2018.

\bibitem{hall2007antennas}
P.~S. Hall, Y.~Hao, Y.~I. Nechayev, A.~Alomainy, C.~C. Constantinou, C.~Parini,
  M.~R. Kamarudin, T.~Z. Salim, D.~T. Hee, R.~Dubrovka \emph{et~al.},
  ``Antennas and propagation for on-body communication systems,'' \emph{IEEE
  Antennas and Propagation Magazine}, vol.~49, no.~3, pp. 41--58, 2007.

\end{thebibliography}

\end{document}